\documentstyle[12pt]{article}
\newcommand{\be}{\begin{equation}}
\newcommand{\ee}{\end{equation}}
\newcommand{\bea}{\begin{eqnarray}}
\newcommand{\eea}{\end{eqnarray}}

\begin{document}
\rightline{BARI-TH/95-214}
\vskip 2cm
{\vbox{\centerline{\bf HEAVY-TO-LIGHT MESON TRANSITIONS IN QCD}}
\bigskip
\bigskip
\centerline{Pietro Colangelo}
\smallskip
\centerline{\it INFN - Sezione di Bari, Italy}
\bigskip
\bigskip
\baselineskip 18pt
\centerline{Abstract}
\bigskip
\noindent
I discuss QCD sum rules determinations of the form factors governing
the decay $B \to \pi (\rho) \ell \nu$.
For some of these form factors the computed dependence on the momentum
transferred does not agree with the expectation from the nearest pole dominance
hypothesis.
Relations are observed among the form factors, that seem to be compatible with
equations recently derived by B.Stech.
The measurement of a number of color suppressed nonleptonic B decay rates could
shed light on the accuracy of the calculation of these form factors and on
the factorization approximation.\\

\bigskip
\vskip 3cm
\centerline{Talk presented at the}
\centerline{6th International Symposium on Heavy Flavour Physics}
\centerline{Pisa, Italy, June 6 - 10, 1995}
\newpage

\setcounter{page}{1}

\noindent
\subsection*{1. Form factors of heavy-to-light meson transitions}
The exclusive semileptonic $B$ decays to $\pi$ and $\rho$ play a prime role in
the measurement of $V_{ub}$. Let us consider, for example, the spectrum of
${\bar {B^0}} \to \pi^+ \mu^- {\bar\nu}$:
\begin{equation}
{d \Gamma ({\bar {B^0}} \to \pi^+ \mu^- {\bar\nu})\over d q^2}=
{G_F^2 \over 24 \pi^3} |V_{ub}|^2 |F_1^{B\pi}(q^2)|^2
|\vec p'_\pi(q^2)|^3 \label{dg}
\end{equation}
where $\vec p'_\pi(q^2)$ is the pion three-momentum (at fixed $q^2$) in the
$B$ meson rest frame. It is clear that a measurement of
$d \Gamma\over d q^2$ would provide us with $V_{ub}$ once the form factor
$F_1^{B\pi}(q^2)$, defined by the hadronic matrix element ($P=\pi$):
\begin{equation}
< P (p^\prime) | \bar q \gamma_\mu b | B(p)> =
 (p + p^\prime)_\mu F_1 (q^2) +
{ M^2_B - M^2_P \over q^2 }
[F_0(q^2) - F_1(q^2)] q_\mu\hskip 2pt \label{sp0}
\end{equation}
\noindent
($q=p-p'$, $F_0(0)=F_1(0)$) is known in the whole accessible range of $q^2$:
$0\le q^2\le q^2_{max}=(M_B-M_\pi)^2$.
An equation analogous to (\ref{dg}) holds for
${\bar {B^0}} \to \rho^+ \mu^- {\bar\nu}$ in terms of the form factors
$V$ and $A_i$ defined by the matrix element ($P^*=\rho$):
\begin{eqnarray}
 <P^* (p^\prime, \eta) | \bar q \gamma_\mu (1- \gamma_5) b
| B(p)>  &=&
 \epsilon_{\mu \nu \rho \sigma}
\eta^{* \nu} p^\rho p^{\prime \sigma} { 2 V(q^2) \over M_B + M_{P^*}}
 -  i  (M_{B} + M_{P^*}) A_1 (q^2) \eta_{\mu}^{\ast} \;
\nonumber \\
 + i   {A_2 (q^2)
\over M_{B} + M_{P^*}} (\eta^{\ast}\cdot p) (p + p^\prime)_\mu
&+& i \;(\eta^{\ast}\cdot p) {2 M_{P^*} \over q^2} q_\mu
\big(A_3 (q^2) - A_0 (q^2)\big)
\label{sp1}
\end{eqnarray}
\noindent ($A_3(0)=A_0(0)$ and
$A_3 (q^2)\;=\;{M_{B}+M_{P^*} \over 2 M_{P^*}} A_1 (q^2) -
{M_{B}-M_{P^*} \over 2 M_{P^*}} A_2 (q^2)\; \;$).

Model independent relations can be derived for $F_i$, $V$ and $A_i$ in the
infinite heavy quark mass limit at the point of
zero recoil ($q_{max}^2$) where $\pi$ and $\rho$ are at rest in the $B-$meson
rest frame \cite{IW1}.
For example, when $m_b \to \infty$ the following relations can be
worked out for $F_1$ and $F_0$:
\begin{eqnarray}
F_1^{B\pi}(q_{max}^2)& \simeq &{g \over 2 f_\pi} {\hat F}
{\sqrt {M_B} \over \Delta +
M_\pi} \label{f1}\\
F_0^{B\pi}(q_{max}^2)& \simeq & - {\hat F \over f_\pi } {1 \over \sqrt {M_B}}
+ {\cal O} (M_\pi) \; ; \label{f0}
\end{eqnarray}
eq.(\ref{f1}) describes the dominance of the $B^*$ pole for $F_1$ at zero
recoil ( in the limit $m_b \to \infty$
$\hat F$ is related to $f_B$ and $f_{B^*}$, $g$ is the rescaled
$B^*B\pi$ strong coupling, and $\Delta=M_{B^*} - M_B$); eq.(\ref{f0}) is
the Callan-Treiman relation valid in the chiral limit.

The above scaling relations, that could be used, e.g., to relate
$B \to \pi \ell \nu$ to $D \to \pi \ell \nu$ at zero recoil, are not sufficient
to describe the form factors in the physical range of
transferred momentum; therefore,  a dynamical calculation
 based on QCD is required for $F_i(q^2)$, $V(q^2)$ and $A_i(q^2)$.

The method of QCD sum rules \cite{SVZ}
is a fully relativistic field-theoretical approach
incorporating fundamental features of QCD, such as perturbative asymptotic
freedom and nonperturbative quark and gluon condensation.
This method  allows us,  by analyzing three-point correlators of
quark currents, to compute the form factors from zero
to quite large values of $q^2$; in
this respect, the method complements lattice QCD, where B meson form
factors, extrapolated from the charm quark mass, are
computed in the region near $q^2_{max}$
\cite{APE,UKQCD,Gu}.

Several QCD sum rules calculations of $F_1^{B \pi}$ can be found in the
literature \cite{Ball}; a calculation of both $F_1$ and $F_0$ has been
performed in ref.\cite{Col1} in the limit $m_b \to \infty$. The result
for $F_1^{B \pi}(q^2)$, depicted in
fig.1 and common to other QCD sum rules analyses,
 supports the simple pole model:
$F_1^{B \pi}(q^2)= {[0.24 \pm 0.02] /( 1- {q^2 \over M^2_{B^*}})}$;
on the other hand $F_0^{B \pi}(q^2)$ increases slowly with $q^2$.
The feature of $F_0$ of
being nearly independent of $q^2$ has been confirmed by a
calculation, at finite $m_b$, in the channel $B \to K$ (fig.2a)
\cite{Col2}.
\vskip 6.truecm\noindent
\includegraphics{f0_bp.ps}
\par
\centerline{Fig.1: Form factors $F_1^{B   \pi}$
(continuous line) and $F_0^{B  \pi}$ (dashed line).}

The computed $q^2$ dependence of $F_0^{B \pi}(q^2)$
must be compared with the
expectation based on the hypothesis of
the dominance of the nearest singularity in the $t-$
channel, assumed in a number of models \cite{mod}:
the nearest pole contributing to
$F_0^{B \pi (B K)}$ is the $0^+$ $b \bar u$  ($b \bar s$) state
with mass in a range near $6 \; GeV$ (in the BSW model the
value $M_{(b {\bar s})}(0^+)=5.89 \; GeV$ is used);
on the other hand, a fit of the obtained
$F_0^{B \pi}(q^2)$ and
$F_0^{B K}(q^2)$ to a simple pole
formula can be performed provided that $M_P \ge 7 - 7.5 \; GeV$.

This different $q^2$ behavior of
$F_1^{B \pi}(q^2)$ and $F_0^{B \pi}(q^2)$
has been observed also in lattice QCD \cite{UKQCD, Gu}. Moreover, a
different functional dependence is expected if one considers that the
scaling laws with the heavy mass in eqs.(\ref{f1},\ref{f0}) are compatible
with the relation $F_1(0)=F_0(0)$ if the $q^2$ dependence is, e.g., of the
type:
$F_i(q^2)= {F_i(0)/ (1- {q^2 \over M^2_i})^{n_i}}$, with
$n_1=n_0+1$.

Deviations from the single pole model have been observed also for the
axial form factors
$A_1^{B \rho}$ and $A_2^{B \rho}$, that turn out to be rather flat in $q^2$
(see the first article in ref.\cite{Ball}); on the other hand,
$V^{B \rho}$ can be fitted with a polar formula, the pole given by $B^*$
\footnote {
A steeper increase of $V$ compared to $A_1$ is obtained
also in ref.\cite{Ali},
but the slopes are different with respect to Ball's results in \cite{Ball}.}.
As for the last form factor in eq.(\ref{sp1}), $A_0$,
the calculation both in the channels $B \to \rho$ and $B \to K^*$
\cite{Col2,Col3} shows that it also increases like a pole, with the
pole mass compatible with the mass of $B$ (or $B_s$) as expected by the
nearest-resonance dominance hypothesis (fig.2b). Interesting enough,
 the relation $A_0(0) \simeq F_0(0)$ is obtained.

To summarize the results from QCD Sum rules analyses, the following scenario
emerges for the transitions $B \to \pi, \rho$ ( $B \to K, K^*$):
$F_1$, $V$ and $A_0$ following a polar dependence in $q^2$,
$F_0$, $A_1$ and $A_2$ rather flat in $q^2$,
$A_0(0)=F_0(0)=F_1(0)$.
It is worth reminding that such results are obtained  after an involved
analytic and numerical analysis, independent for each one of the above
form factors.
\vskip 6.5 truecm\par
\includegraphics{f0_bk.ps}
\includegraphics{a0_bks.ps}
\par
\vskip 0.5 cm
\centerline{ Fig.2: Form factors $F_0^{B   K}$ (a) and
$A_0^{B  K^*}$ (b).}

One could wonder whether
QCD sum rules results suggest the existence of relations
among the form factors governing the transitions of heavy mesons to light
mesons. For semileptonic decays where both the initial
and the final meson contains one heavy quark, such relations can be derived
in the limit $m_Q \to \infty$:
they connect the six form factors as in
(\ref{sp0},\ref{sp1}) to the Isgur-Wise function \cite{IW} incorporating
the nonperturbative dynamics of the light degrees of freedom.
It is intriguing that relations among heavy-to-light form factors have been
obtained in a constituent quark model by B.Stech \cite{Stech},
assuming that the spectator particle retains its
momentum and spin before the hadronization, and that
in the rest frame of the hadron the constituent quarks
have the off-shell energy close to the constituent mass.
Under these hypotheses the following equations can be written for
$B \to \pi, \rho$ form factors
\footnote{ A dependence of $F_1$ on the mass of the final particle has
to be taken into account since there is no spin symmetry in the
final state.}:
\begin{eqnarray}
&&F_0(q^2)=\left(1-\frac{q^2}{m_B^2-m^2_\pi}\right)
F_1(q^2) \label{r1} \\
&&V(q^2)=\left(1+\frac{m_\rho}{m_B}\right)
F_1(q^2) \label{r2}  \\
&&A_1(q^2)=\frac{1+\frac{m^2_\rho}{m^2_B}}
{1+\frac{m_\rho}{m_B}}\left
(1-\frac{q^2}{m^2_B+m^2_\rho}\right)F_1(q^2) \label{r3} \\
&&A_2(q^2)=\left(1+\frac{m_\rho}{m_B}\right)\left(
1-\frac{2m_\rho/(m_B+m_\rho)}{1-q^2/(m_B+m_\rho)^2}\right)
F_1(q^2) \label{r4} \\
&&A_0(q^2)=F_1(q^2). \label{r5}
\end{eqnarray}
The above relations are very similar to the relations holding
for heavy-to-heavy transitions, e.g. $B \to D, D^*$.
QCD sum rules results seem to confirm them.
As a matter of fact, for a polar dependence of
$F_1(q^2)$ the above relations suggest that both
$F_0$ and $A_1$ should be nearly constant in $q^2$, and that
$A_0$ should be equal to $F_1$.

Of course, more work is needed to put
equations (\ref{r1}-\ref{r5}) on the same theoretical grounds of
 the relations among the form factors of heavy-to-heavy transitions.

\subsection*{2. Tests of factorization for color suppressed B decays}

Semileptonic form factors are useful not only to predict
semileptonic BR's, but
also to compute nonleptonic two-body decay rates if the factorization
approximation is adopted. In particular, for color suppressed transitions
$B \to K^{(*)} \; J/\Psi$ and
$B \to K^{(*)} \; \eta_c$,  only the heavy-to-light form factors are needed.
The decays
$B \to K^{(*)} \; J/\Psi $ have been analyzed in
\cite{Gourdin} to constrain  the semileptonic
$B \to K^*, K $ form factors using data on the
longitudinal polarization of the final
particles in the decay $B \to K^* \; J/\Psi$:
$\rho_L=\Gamma(B \to K^* \; J/\Psi)_{LL}/\Gamma(B \to K^* \; J/\Psi)
=0.84 \pm 0.06\pm 0.08$, and on the ratio
$R_{J/\Psi}=\Gamma(B \to K^* \; J/\Psi)/\Gamma(B \to K \; J/\Psi)
=1.64 \pm 0.34$ \cite{Hon}.
In the same spirit, the decays $B \to K^{(*)} \; \eta_c $ are interesting
since they could help
in testing the factorization scheme and the accuracy of the computed
hadronic quantities \cite{Col2}.

To predict the decay rates $B \to K^{(*)} \; \eta_c $, besides
$F_0^{B  K}(q^2)$ and $A_0^{B  K^*}(q^2)$ we need  the leptonic
constant $f_{\eta_c}$. Together with $f_{\eta'_{c}}$, $f_{\eta_c}$
can be obtained by QCD sum rules considering the two-point function
\begin{eqnarray}
\psi_{5}(q) = i \int d^{4}x \; e^{iqx}\; < 0| T (\partial^{\mu} A_{\mu} (x)
	\partial^{\nu} A_{\nu}^{\dagger} (0)) |0 > \;
\end{eqnarray}
($\partial ^{\mu} A_{\mu} (x) = 2 m_{c} : \bar{c}(x) i \gamma_{5}c(x):$)
that is known in perturbative QCD to two-loop
order, including also  the leading $D=4$ non-perturbative
term in the Operator Product Expansion.
Exploiting two different types
of QCD sum rules, viz. Hilbert transforms at $Q^{2} = 0$, and Laplace
transforms, we get \cite{Col2}:
\begin{equation}
f_{\eta_{c}} \simeq 301 - 326 \; \mbox{MeV},\;\;\;\;\;\;\;\;
f_{\eta'_{c}} \simeq 231 - 255 \; \mbox{MeV},\;
\end{equation}
\begin{equation}
f_{\eta_{c}} \simeq 292 - 310 \; \mbox{MeV},\;\;\;\;\;\;\;\;
f_{\eta'_{c}} \simeq 247 - 269 \; \mbox{MeV}\;,
\end{equation}
respectively. These results have been obtained by varying the parameters
in the ranges
dictated by the gluon condensate and quark-mass analyses, and using
$m_{c} = 1.46 \pm 0.07 \; \mbox{GeV}$,
$\Lambda_{QCD} = 200 - 300 \; \mbox{MeV}$, with the constraint that
$M_{\eta_{c}}$ and $M_{\eta'_{c}}=3595\pm5\;MeV$ are
correctly reproduced by the sum rules.
Combining the predictions from the Hilbert and Laplace  method we obtain:
\begin{equation}
f_{\eta_{c}} = 309 \pm  17 \; \mbox{MeV},\;\;\;\;\;\;\;\;
f_{\eta'_{c}} = 250 \pm 19 \; \mbox{MeV},\;\;\;\;\;\;\;\;
\frac{f_{\eta'_{c}}} {f_{\eta_{c}}} = 0.8 \pm 0.1 \; ,
\label{pred}
\end{equation}
\begin{equation}
{f_{\eta_c} \over f_{J/\psi}} = 0.81 \pm 0.05,\;\;\;\;\;\;\;\;
{f_{\eta^\prime_c} \over f_{\Psi^\prime}} = 0.88 \pm 0.08 .
\label{pred1}
\end{equation}
In (\ref{pred1}) the experimental values: $f_{J/\Psi}=384 \pm 14 \; MeV$ and
$f_{\Psi^\prime}=282 \pm 14 \; MeV$ have been used.
In the constituent quark model the leptonic constants of the charmonium system
can be expressed in terms of the $c \bar c$ wave function at the origin
$\Psi(0)$:
\begin{equation}
f_{\eta_c}^2 = 48 {m_c^2 \over M_{\eta_c}^3} |\Psi(0)|^2\; , \;\;\;\;\;\;\;
f_{J/\Psi}^2 = 12 {1 \over M_{J/\Psi}} |\Psi(0)|^2 \; ;
\end{equation}
therefore, the ratio $f_{\eta_c}/f_{J/\Psi}$ can be predicted in terms
of the meson masses and of the charm quark mass:
\begin{equation}
{f_{\eta_c} \over f_{J/\Psi}} =
2 m_c \Big( {M_{J/\Psi} \over M_{\eta_c}^3}\Big)^{1\over 2} = 0.97 \pm 0.03 \;;
\label{ratio}
\end{equation}
the deviations from the outcome of QCD sum rules,
at the level of $15- 20 \%$ for
$\eta_c$, $J/\Psi$ and $5 - 8 \%$ for
the radial excitations $\eta^\prime_c$, $\Psi^\prime$, can be
attributed to relativistic and radiative corrections to the constituent quark
model formula.

Tests of factorization can be performed by analyzing ratios of
decay widths, such as  $B \to K^{(*)} \eta_c$ and
 $B \to K^{(*)} \eta_c^\prime$, where the dependence on the Wilson
coefficients in the effective hamiltonian governing the decays,
and on other weak parameters drops out.
Let us consider, e.g., the ratio:
\begin{equation}
{\tilde R}_K = { \Gamma(B^- \to K^- \eta^\prime_c) \over
\Gamma(B^- \to K^- \eta_c)} = \; 0.771 \;
\big({f_{\eta^\prime_c} \over f_{\eta_c}}\big)^2 \;
\big({F_0(M^2_{\eta^\prime_c}) \over
F_0(M^2_{\eta_c})} \big)^2 = 0.60 \pm 0.15  \; .
\end{equation}
The interesting point is that, because of the flat shape of
$F_0(q^2)$ (fig.2a), ${\tilde R}_K$ mainly depends
on the ratio of the leptonic constants, so that
in factorization approximation a measurement of ${\tilde R}_K$
would provide us with interesting information on
${f_{\eta^\prime_c} \over f_{\eta_c}}$, and complement our knowledge of the
${c\bar c}$ wavefunction.
The analogous ratio for the decays into $K^*$ is given by
\begin{equation}
{\tilde R}_{K^*} = { \Gamma(B^- \to K^{*-} \eta^\prime_c) \over
\Gamma(B^- \to K^{*-} \eta_c)} =
0.381 \; \big({f_{\eta^\prime_c} \over f_{\eta_c}}\big)^2 \;
\big({A_0(M^2_{\eta^\prime_c})
\over A_0(M^2_{\eta_c})} \big)^2 =
0.381 \; \big({f_{\eta^\prime_c} \over f_{\eta_c}}\big)^2 \; (1.4 \pm 0.2)^2
\; .\label{polar}\end{equation}
\noindent
Here, the ratio of the form factors deviates from unity due to the
$q^2$-dependence of $A_0$ (fig.2b). The prediction from (\ref{polar})
would be:
${\tilde R}_{K^*} = 0.45 \pm 0.16$. Moreover,
the quantity $\sqrt{ {\tilde R}_{K^*}/{\tilde R}_{K}}$
is sensitive to the $q^2$-dependence of the ratio $A_0/F_0$:
\begin{equation}
1.42 \; \sqrt{  { {\tilde R}_{K^*}\over {\tilde R}_{K} } }=
\Big( { A_0(M^2_{\eta^\prime_c})/F_0(M^2_{\eta^\prime_c}) \over
A_0(M^2_{\eta_c}) /F_0(M^2_{\eta_c}) } \Big) \;.
\end{equation}
We also get:
\begin{equation}
{R}_{\eta_c} = { \Gamma(B^- \to K^{*-} \eta_c) \over
\Gamma(B^- \to K^- \eta_c)}= 0.373 \;
\big({A_0(M^2_{\eta_c})
\over F_0(M^2_{\eta_c})} \big)^2 =
0.73 \pm 0.13
\end{equation}
\noindent
and ${ R}_{\eta^\prime_c}=0.56 \pm 0.12$ for the analogous  ratio
${ R}_{\eta^\prime_c}$.
Finally, the ratio:
\begin{equation}
R_K= { \Gamma(B^- \to K^{-} \eta_c) \over
\Gamma(B^- \to K^{-} J/\Psi)} =
2.519 \; \big({f_{\eta_c} \over f_{J/\Psi}}\big)^2 \;
\big({F_0(M^2_{\eta_c})
\over F_1(M^2_{J/\Psi})} \big)^2 \;
\end{equation}
\noindent can be predicted using
the simple pole model for $F_1^{B  K}$.
We obtain:
$R_K=0.94 \pm 0.25$, and, for
$R^\prime_K= { \Gamma(B^- \to K^{-} \eta^\prime_c) \over
\Gamma(B^- \to K^{-} \Psi^\prime)}$:
$R^\prime_K= 1.61 \pm 0.53$.
Using the CLEOII measurements \cite{Hon}:
${\cal B}(B^- \to K^{-} J/\Psi) = (0.11 \pm 0.01 \pm 0.01)\times 10^{-2}$
and
${\cal B}(B^- \to K^{-} \Psi^\prime) = (0.06 \pm 0.02 \pm 0.01)\times 10^{-2}$
we expect:
${\cal B}(B^- \to K^{-} \eta_c) = (0.11 \pm 0.03)\times 10^{-2}$
and
${\cal B}(B^- \to K^{-} \eta_c^\prime) = (0.10 \pm 0.05)\times 10^{-2}$,
that should be within reach of present experimental facilities.
The measurement of some of the above decay rates
could shed more light on the problem of factorization,
which is a basic assumption in the present analysis of heavy meson nonleptonic
decays.

\noindent {\bf Acknowledgments. }
It is a pleasure to thank  F.De~Fazio, C.A.Dominguez, G.Nardulli, N.Paver and
P.Santorelli for their collaboration on the subjects discussed here.

\end{document}